\documentclass[aps,superscriptaddress,twocolumn,showpacs,preprintnumbers,amsmath,amssymb]{revtex4}
\usepackage[dvips]{graphicx}
\usepackage{eucal} \usepackage{bm} 
\usepackage{float}
\usepackage{color}
\newcommand{\pT}{p_{\scriptscriptstyle T}}

\newcommand{\lsim}{\mbox{\raisebox{-0.6ex}{$\stackrel{<}{\sim}$}}\:}

\begin{document}
\title{
Hadron-string cascade versus hydrodynamics
in Cu + Cu collisions
at $\sqrt{s_{NN}}=200$ GeV
}

\medskip

\author{T. Hirano}
\affiliation{
Department of Physics, Columbia University, 
New York, NY 10027}
\author{M. Isse}
\affiliation{
 Division of Physics,
 Graduate School of Science, Hokkaido University,
Sapporo, Hokkaido 060-0810, Japan }

\author{Y. Nara}
\affiliation{
Institut f\"ur Theoretische Physik,
Johann Wolfgang Goethe-Universit\"at,
Max-von-Laue-Str. 1, 60438  Frankfurt am Main, Germany
}
\author{A. Ohnishi}
\affiliation{
 Division of Physics,
 Graduate School of Science, Hokkaido University,
Sapporo, Hokkaido 060-0810, Japan }

\author{K. Yoshino}
\affiliation{
 Division of Physics,
 Graduate School of Science, Hokkaido University,
Sapporo, Hokkaido 060-0810, Japan }

\begin{abstract}
Single particle spectra as well as elliptic flow
in Cu+Cu collisions at $\sqrt{s_{NN}}=200$ GeV are investigated
within a hadronic cascade model and an ideal hydrodynamic model.
Pseudorapidity distribution and transverse momentum spectra
for charged hadrons are surprisingly comparable between these two models.
However, a large deviation is predicted for the elliptic flow.
The forthcoming experimental data will
clarify the transport and thermalization aspects of matter
produced in Cu+Cu collisions.
\end{abstract}

\maketitle

One of the primary current interests 
in the Relativistic Heavy Ion Collider (RHIC) experiments
is to explore the properties of
QCD matter far from stable nuclei, especially the confirmation of
the deconfined and thermalized matter,
\textit{i.e.} the quark gluon plasma (QGP),
which has been predicted from the lattice QCD calculations~\cite{lattice}.
While high and medium $\pT$ observables
such as the parton energy loss~\cite{e-loss}
and coalescence behavior of hadron elliptic flows~\cite{Fries:2004ej}
are generally believed to give strong evidences of
high dense
matter formation,
hadrons at these momenta are not necessarily formed from thermalized matter.
Therefore, low $\pT$ observables are also important to confirm 
whether \textit{equilibrium} is achieved or not.

Elliptic flow \cite{Ollitrault:1992bk} is one of the promising observables
to study the degree of thermalization for
 QCD matter produced in heavy ion collisions
since it is believed to be sensitive to the properties of the
matter at initial stages and the collision 
geometry~\cite{Heiselberg:1998es,Zhang:1999rs}.
Indeed, incident energy as well as impact parameter dependences
of elliptic flow have been investigated extensively.
Elliptic flow, \textit{i.e.} the momentum anisotropy
with respect to the reaction plane $v_2=\langle \cos(2\phi) \rangle$,
has been measured in a wide energy range from
GSI-SIS ($E_\mathrm{inc}\lesssim 1 A$ GeV)~\cite{sis},
BNL-AGS ($E_\mathrm{inc}=2-11 A$ GeV)~\cite{ags},
to CERN-SPS ($E_\mathrm{inc} = 40-158 A$ GeV)~\cite{NA49},
in addition to BNL-RHIC~\cite{RHICflow}.
Measured collective flows are well reproduced by 
nuclear transport models assuming the 
momentum dependent 
nuclear mean-field
at SIS to AGS ($E_\mathrm{inc}\simeq 0.2 - 11 A$GeV)~\cite{da02,Giessen}
and SPS ($E_\mathrm{inc}=40, 158 A$GeV)~\cite{isse05} energies,
whereas elliptic flow at RHIC at mid-rapidity is underestimated
in nonequilibrium transport models
which do not include explicit partonic interactions
~\cite{Bleicher2000,Cassing03,Cassing04}.
It is also reported that hadronic models explain
elliptic flow only at low transverse momentum $p_T\lesssim 1$ GeV/$c$
at RHIC~\cite{Burau:2004ev}.
Partonic interactions followed by quark coalescence
hadronization mechanism
are proposed in Ref.~\cite{lin02}
to account for the experimental data on elliptic flow. 
Note, however, that hadronic cascade models reproduce
elliptic flow in forward/backward rapidity regions at 
RHIC~\cite{Stoecker:2004qu}.

On the other hand, in Au+Au collisions at RHIC energies
the magnitude of $v_2$ and its transverse momentum $p_T$
and mass $m$ dependences are close to predictions
based on ideal and non-dissipative hydrodynamics
simulations around midrapidity ($\mid \eta \mid \lsim 1$),
in the low transverse momentum
region ($p_T \lsim 1$ GeV/$c$), and up to semicentral
collisions ($b \lsim 5$ fm)
\cite{Kolb:2000fh,Hirano:2001eu}.
This is one of the main results
which leads to a recent announcement of the discovery
of perfect fluidity at RHIC \cite{BNL}.
(See Ref.~\cite{HiranoGyulassy} for recent reinterpretation
of the RHIC data based on current hydrodynamic results.)
Despite the apparent success near midrapidity at RHIC,
ideal hydrodynamics
overestimates the data
at lower incident energies (SIS, AGS and SPS)
as well as in forward/backward rapidity regions at RHIC
probably due to the lack of dissipative effects.

We study Cu+Cu collisions at RHIC in the present work, 
which is a complementary study of
elliptic flow in Au+Au collisions.
The particle density and the size of the system
are smaller in Cu+Cu collisions than in Au+Au collisions.
So the reasonable agreement of hydrodynamic results
with Au+Au data
may be spoiled in Cu+Cu collisions
and a non-equilibrium hadronic
description can be relatively important
even at RHIC energies.
Therefore we employ both a hadronic transport model JAM
and a hydrodynamic model
to
make predictions for elliptic flow
in Cu+Cu collisions.
Below we briefly summarize hadron-string cascade JAM~\cite{jam}
and a hydrodynamic model \cite{Hirano:2004rs} adopted in this paper.

A hadronic transport model JAM simulates nuclear collisions
by the individual hadron-hadron collisions.
Soft hadron productions in hadron-hadron scattering
are modeled by the resonance and color string excitations.
Hard partonic scattering is also included in line with HIJING~\cite{hijing}.
Color strings decay into hadrons
after their formation time $(\tau \sim 1$ fm/$c$)
according to the Lund string model PYTHIA~\cite{Sjostrand}.
Hadrons within their formation time can scatter 
with other hadrons assuming the additive quark cross section. 
This simulates constituent quark collisions effectively
which is known to be important at SPS energies~\cite{urqmd}.
Therefore, matter initially created in collisions is represented by the
many strings at RHIC, which means that there is no QGP in the model.

Default parameters in JAM are adopted in this work
except for a little wider $\pT$ width in the string decay
and a larger partonic minimum $\pT$ ($p_0=2.7 \mbox{GeV}/c$)
to fit charged hadron $\pT$ spectrum in $pp$ collisions at
$\sqrt{s_{NN}}=200$ GeV.
In addition to hadron-hadron collisions,
nuclear mean field is incorporated in JAM and
its effects are known to be important at AGS and SPS energies~\cite{isse05},
but mean field is not expected to play major roles at RHIC.
We have thus neglected nuclear mean field in this work.
The detailed description of JAM can be found in Ref.~\cite{jam}.

Two of the authors (T.H. and Y.N.)
have already developed another dynamical framework
to describe three important aspects of relativistic heavy ion
collisions~\cite{Hirano:2004rs},
namely color glass condensate (CGC) for collisions
of two nuclei~\cite{MV,Iancu:2003xm,KLN}, 
hydrodynamics for space-time evolution
of thermalized matter~\cite{hydroreview},
and jet quenching for high $p_T$ non-thermalized partons~\cite{GLV}.
Along the line of these works, we use the same model in this study.
However, our aim is to study the bulk properties
of matter produced in Cu+Cu collisions.
In this paper, we neither include jet components in this model
nor discuss jet quenching, unlike a series of 
the previous work~\cite{Hirano:2002sc}.
So hydrodynamic results to be presented below
include purely boosted thermal components
without any semi-hard components.

In Ref.~\cite{Hirano:2004rs},
a systematic hydrodynamic analysis
in Au+Au collisions at $\sqrt{s_{NN}}=200$ GeV
was performed by using 
initial conditions taken from the CGC
picture for the colliding nuclei.
In the conventional hydrodynamic calculations,
one chooses initial condition for hydrodynamic equations
and thermal freezeout temperature $T^{\mathrm{th}}$
so as to reproduce the \textit{observed} particle spectra, such as
(pseudo)rapidity distribution and transverse momentum
distribution.
So it is believed that
hydrodynamics has a less predictive power compared with
cascade models.
However, if the initial particle production
at high collisional energies
is supposed to be universal as described by the CGC,
hydrodynamics with CGC initial conditions
can \textit{predict} particle spectra.
Here, we employ the IC-$n$,
\textit{i.e.} a prescription that
the number density produced
in a CGC collision is matched to the hydrodynamic
initial condition \cite{Hirano:2004rs},
to obtain the initial distribution of
thermodynamic variables
at the initial time $\tau_0$.
Once the initial condition is obtained,
one solves hydrodynamic equation $\partial_\mu T^{\mu\nu}=0$
in the three-dimensional
Bjorken coordinate $(\tau, \eta_s, x, y)$ \cite{Hirano:2001eu}.
Here we neglect a dissipative effect and a 
finite (but probably tiny)
baryon density.
Assuming $N_{c} = N_{f}=3$ massless partonic gas,
an ideal gas equation of state (EOS)
 with a bag constant $B^{1/4}=247$ MeV is 
employed in the QGP phase ($T>T_c=170$ MeV).
We use a hadronic resonance gas model
with all hadrons up to $\Delta(1232)$
mass for later stages ($T<T_c$) of collisions. 
We take into account chemical freezeout
separated from thermal freezeout \cite{Hirano:2002ds}
as required to obtain sufficient yields for heavier particles. 
Specifically, we assume that 
chemical freezeout temperature $T^{\mathrm{ch}}=170$ MeV and
kinetic freezeout temperature $T^{\mathrm{th}}=100$ MeV.
Note that the slope of $p_T$ spectra becomes
insensitive to $T^{\mathrm{th}}$
(while $v_2$ becomes sensitive)
when chemical freezeout is taken into account \cite{Hirano:2002ds}.
In the calculation of $v_2(\eta)$, we also use 
$T^{\mathrm{th}}=160$ MeV
for comparison.
If the strongly coupled QGP (sQGP) core expands as a perfect fluid
and the hadronic corona does as a highly dissipative gas
as suggested in Ref.~\cite{HiranoGyulassy},
the resultant $v_2(\eta)$ and $v_2(p_T)$ are
expected to be frozen
after hadronization \cite{Teaney:2000cw,Teaney:2002aj}
due to the strong viscous effect.
Moreover, $T^{\mathrm{th}}$ should be higher for
a smaller size of the system \cite{Hung:1997du} as observed
in the centrality dependence of the $p_T$ spectra \cite{Adams:2003xp}.
So the freezeout picture in Cu+Cu collisions can be different from
that in Au+Au collisions.
In the following predictions for hydrodynamic elliptic flow,
we show the results 
for $T^{\mathrm{th}} = $ 100 and 160 MeV.
For further details of the hydrodynamic model used in this work,
see Refs.~\cite{Hirano:2004rs,Hirano:2002sc,Hirano:2002ds}.

\begin{figure}[tb]
\begin{center}
\includegraphics[width=9cm,clip]{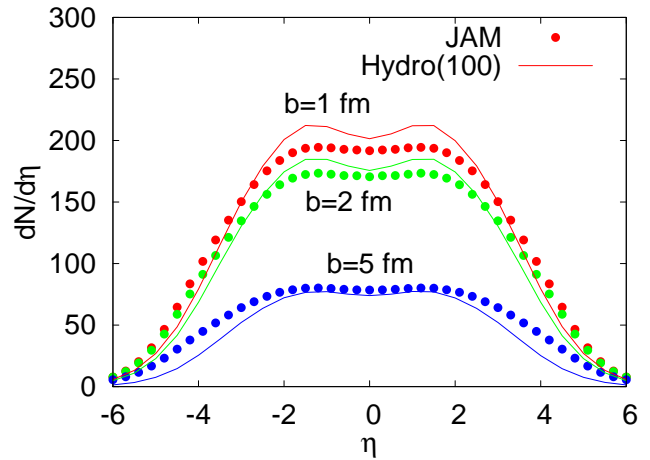}
\caption{(Color online) Pseudorapidity distributions
for charged hadrons in Cu+Cu collisions
at $\sqrt{s_{NN}} = 200$ GeV
for impact parameters $b=$ 1, 2, and 5 fm.
Circles correspond to the result of JAM.
Lines denote the 
results of hydrodynamics for $T^{\mathrm{th}}=100$ MeV.
}
\label{dndeta}
\end{center}
\end{figure} 
\begin{figure}[tb]
\begin{center}
\includegraphics[width=9cm,clip]{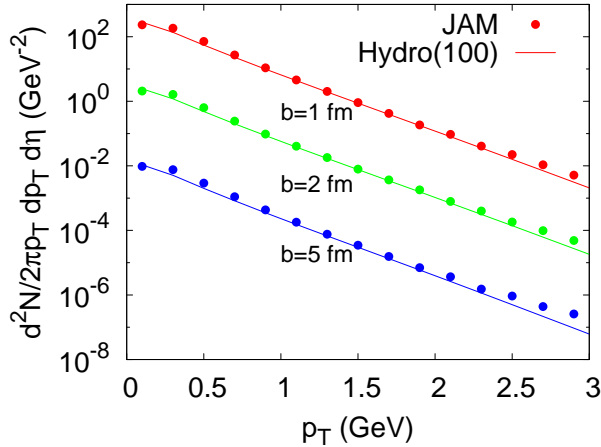}
\caption{(Color online) Comparison of the transverse momentum distributions
for charged hadrons between JAM (circles) and hydrodynamics 
for $T^{\mathrm{th}}=100$ MeV (lines)
at $|\eta|<0.33$ in Cu+Cu collisions at $\sqrt{s_{NN}}=200 $GeV.
The results for $b$=2 fm (5 fm) are scaled by 10$^{-2}$ (10$^{-4}$).
}
\label{dndpt}
\end{center}
\end{figure}

We first compare the bulk single particle spectra
between JAM and hydrodynamics.
We emphasize again that our hydrodynamic results
are insensitive to a choice
of $T^{\mathrm{th}}$ for transverse and rapidity
distributions of charged hadrons.
We show results of the
pseudorapidity distribution $dN/d\eta$
for charged hadrons in Fig.~\ref{dndeta}
at impact parameters $b=1, 2$ and $5$ fm.
It is seen from this figure that
the shape and the magnitude of the distributions
from JAM are almost similar to those from hydrodynamics.

In Fig.~\ref{dndpt},
we compare JAM and hydrodynamic results of
the $\pT$ spectra for charged hadrons
at impact parameters of $b=1, 2$ and 5 fm for $|\eta|<0.33$.
Accidentally, 
these results agree well with each other in transverse momentum
range of $\pT<2$ GeV/$c$. Deviation at higher transverse momentum
is due to the lack of jet components in the hydrodynamic simulations.

At least within our models,
two distinct pictures, \textit{i.e.}
pictures of coherent particle production via CGC
combined with sequential sQGP expansion and 
of transports of secondary hadrons
after hadron-hadron collisions summed up by
an overlap region of colliding nuclei,
 are indistinguishable in the bulk single
hadron distributions in Cu+Cu collisions.
Note that free parameters in the ``CGC+hydro" model has been fixed
by fitting the charged multiplicity in Au+Au collisions at midrapidity.
We also note that parameters in JAM are already fixed
to fit the data in $pp$ collisions at $\sqrt{s_{NN}}=200$ GeV.

\begin{figure}[tb]
\begin{center}
\includegraphics[width=9cm,clip]{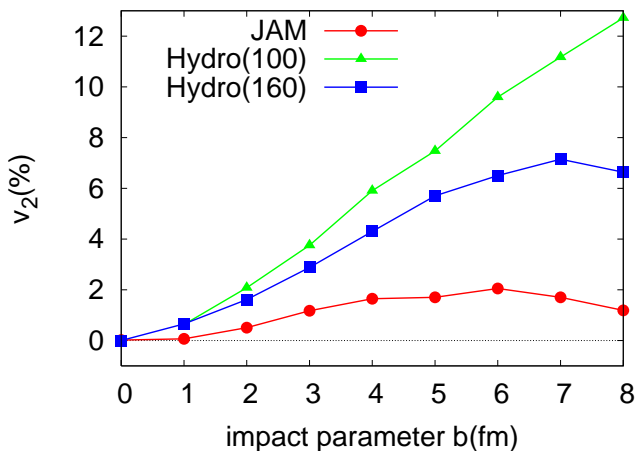}
\caption{(Color online)
Elliptic flow $v_2$ for charged hadrons 
at mid-rapidity as a function of impact parameter 
in Cu+Cu collisions at $\sqrt{s_{NN}}=200 $GeV.
Circles connected by line show the results of JAM.
Triangles and squares with lines show the results of hydrodynamics
with $T^{\mathrm{th}}=$100 MeV and 160 MeV, respectively.
}\label{v2b}
\end{center}
\end{figure} 
\begin{figure}[tb]
\begin{center}
\includegraphics[width=9cm,clip]{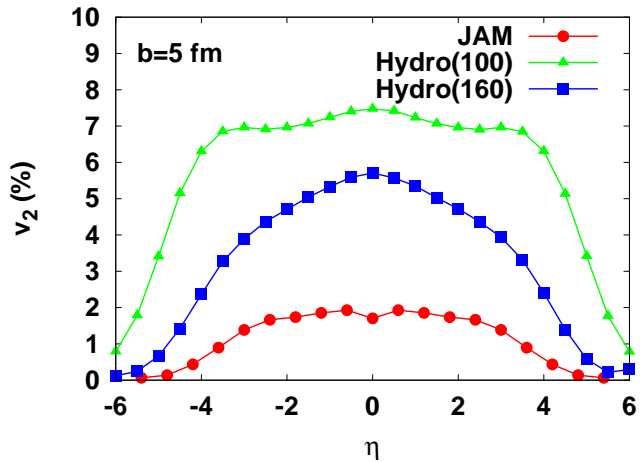}
\caption{(Color online) Elliptic flow $v_2$ for charged hadrons
as a function of pseudorapidity $\eta$ 
in Cu+Cu collisions
at $\sqrt{s_{NN}}=200$ GeV at impact parameter
$b$ = 5 fm.
Circles connected by line show the results of JAM.
Triangles and squares with lines show the results of hydrodynamics
with $T^{\mathrm{th}}=$100 MeV and 160 MeV, respectively.
}
\label{fig:v2eta_JAM_hydro}
\end{center}
\end{figure}

In Fig.~\ref{v2b}, we show the impact parameter $b$ dependence of
the elliptic flow $v_2$ at mid-rapidity for charged hadrons.
In the hydrodynamic calculations, kinetic freezeout temperatures
$T^{\mathrm{th}}=100$ MeV and 160 MeV are chosen.
While single particle spectra
from JAM and hydrodynamics
look very similar,
a clear difference of $v_2(b)$ is seen:
$v_2$ grows almost linearly with $b$ in hydrodynamics,
which is the same as the case in Au+Au collisions,
while we find a peak at around $b=6$ fm in JAM
and that the magnitude is only around 20\% of the hydrodynamic prediction
with $T^\mathrm{th}=100$ MeV.
The two distinct pictures within our approach
appear differently in the centrality dependence of elliptic flow.
Due to the smaller
initial energy density
in Cu+Cu collisions
compared to Au+Au collisions,
the spatial anisotropy is still out-of-plane just after the hadronization
and $v_2$ continues to be generated
even in the late non-viscous hadronic stage
in the ideal hydrodynamic simulation.
The data is expected to be comparable with the result 
for $T^{\mathrm{th}}$ = 160 MeV
if the initial energy density is large ($e_0 \gg 1$ GeV/fm$^3$)
and the equilibration time is small ($\tau_0\sim 1$ fm/$c$)
enough to create the sQGP phase in Cu+Cu collisions.
On the contrary,
one expects that it takes more time to reach equilibrium ($\tau_0 > 1$ fm/$c$)
and that the system may not reach the equilibrated sQGP state
since the system size and the produced particle number are small
compared with those in Au+Au collisions.
In that case, 
the data will be comparable with the result from JAM.

Pseudorapidity dependences of the elliptic flow
from JAM and hydrodynamics are compared with each other
to understand the longitudinal dynamics in Cu+Cu collisions
in Fig.~\ref{fig:v2eta_JAM_hydro}.
In JAM, we find almost flat
behavior of $v_2(\eta)$ around midrapidity ($|\eta|<2$),
where the charged hadron $\eta$ distribution also shows flat behavior.
In JAM, elliptic flow is slowly generated ($t \lesssim 10 \mathrm{fm}$)
as the hadrons are formed from strings
after some formation times. 
In the hydrodynamic calculations, 
we show the results for $T^{\mathrm{th}}=100$ and 160 MeV,
which could be an upper and
a lower limit of the \textit{ideal} hydrodynamic prediction
respectively.
$v_2(\eta)$ for $T^{\mathrm{th}} = 100$ MeV
becomes a trapezoidal shape, which
looks similar to the result in the previous hydrodynamic
study in Au+Au collisions \cite{Hirano:2001eu,Hirano:2002ds}.
$v_2(\eta)$ from ideal hydrodynamics for $T^{\mathrm{th}} = 160$ MeV
is also shown as a possible result for the 
situation \cite{HiranoGyulassy} in which
$v_2$ is generated by the perfect fluid of the sQGP core
and is not generated significantly in the dissipative
hadronic corona like the result from JAM.
Indeed, $v_2(\eta)$ for $T^{\mathrm{th}}=160$ MeV appears to be
a triangle shape which looks similar to the
shape in Au+Au data observed by PHOBOS \cite{phobos02}.

\begin{figure}[tb]
\begin{center}
\includegraphics[width=9cm,clip]{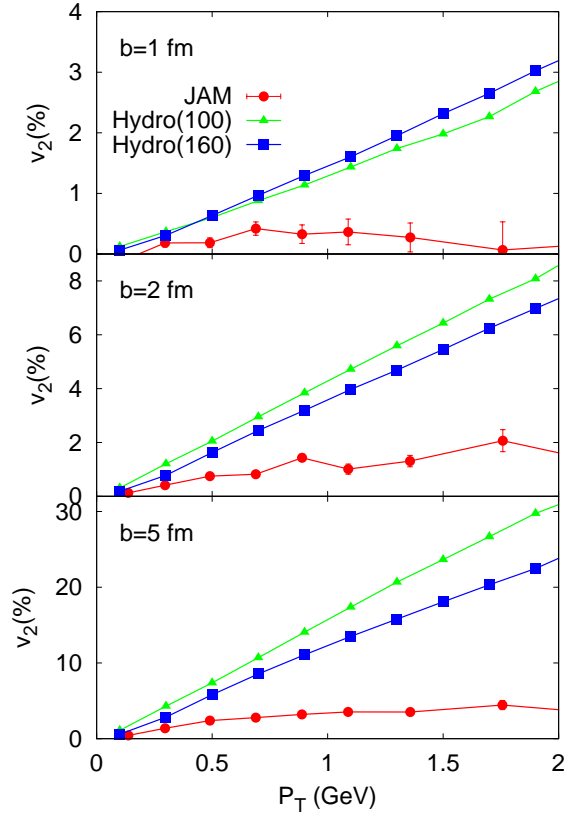}
\caption{(Color online) The calculated elliptic flow $v_2$ of charged hadrons
 as a function of transverse momentum $p_T$
 for Cu+Cu at $\sqrt{s_{NN}}=200$ GeV for different impact parameters,
$b=$ 1 fm (top), 2 fm (middle), and 5 fm (bottom).
Circles connected by line show the results of JAM.
Triangles and squares with lines show the results of hydrodynamics
with $T^{\mathrm{th}}=$100 MeV and 160 MeV, respectively.
}
\label{v2pt}
\end{center}
\end{figure} 

In Fig.~\ref{v2pt}, we compare
transverse momentum $\pT$ dependence of elliptic flow
for charged hadrons.
Hydrodynamic predictions
are of course larger than the ones of JAM.
In JAM, $v_2$ starts to be saturated at around 0.8 GeV,
and the behavior is qualitatively similar to that in Au+Au collisions
and another theoretical prediction
in Cu+Cu collisions \cite{chen05}.
It should be noted that we will also find
a mass dependent saturating behavior of $v_2(\pT)$
when semihard components are combined
with the hydrodynamic components
\cite{Hirano:2004rs}.

In summary, we have investigated low-$\pT$ observables
in a hadron-string cascade model JAM~\cite{jam} and
a hydrodynamical model~\cite{Hirano:2004rs}
in Cu+Cu collisions at $\sqrt{s_{NN}} = 200 $ GeV.
For $dN/d\eta$ and $\pT$-spectra for charged hadrons,
we have obtained good agreement between JAM and hydrodynamics.
However, clear deviations between model predictions are found in
 $v_2$ as a function of centrality,
pseudorapidity $\eta$,
and transverse momentum $\pT$ for charged hadrons.
The lack of elliptic flow in hadronic transport models
compared to the ideal hydrodynamic predictions
is due to the initial particle production being performed
by string decays which only generate a limited amount
of transverse momentum for the produced
particles in conjunction with the formation time
for these hadrons.
In that sense,
the ``EOS'' of hadron-string cascade models
in the very early stage is to be considered as
a super-soft one and cannot generate sufficient pressure
needed for elliptic flow to develop.
In addition, 
even if full thermalization is achieved in the hadron cascade model,
a higher viscosity in the hadron cascade model
would yield a lower
elliptic flow than the ideal non-viscous hydrodynamics
in the hadron phase.
Therefore, in order to interpret correctly the result,
$v_{2,\mathrm{cascade}}<v_{2,\mathrm{hydro}}$,
which has been seen also in SPS energies,
we should study the dissipative effects carefully.
Measurements of
pseudorapidity and transverse momentum dependence of
elliptic flow ($v_2(b)$, $v_2(\eta)$, and $v_2(p_T)$)
in Cu+Cu collisions at RHIC
will provide very important information for
transport aspects of QCD matter in heavy ion collisions.

One of the authors (Y.N.) acknowledges discussions with M. Bleicher.
The work was supported in part by the US-DOE
under Grant No.~DE-FG02-93ER40764 (T.H.),
and by the Grant-in-Aid for Scientific Research No.~1554024~(A.O.)
from the Ministry of Education, Science and Culture, Japan.

\end{document}